\documentclass[reprint, aps, pre, superscriptaddress, showpacs, floatfix]{revtex4-1}
\usepackage{amsmath}
\usepackage{amsthm}
\usepackage{amsfonts}

\usepackage{graphicx}%
\usepackage{chemarr}
\usepackage{dcolumn}%
\usepackage{bm}%
\usepackage{color}

\begin{document}

\begin{abstract}
We investigate the aggregation kinetics of a simulated telechelic polymer gel.  In the hybrid Molecular Dynamics (MD) / Monte Carlo (MC) algorithm, aggregates of associating end groups form and break according to MC rules, while the position of the polymers in space is dictated by MD.  As a result, the aggregate sizes change every time step.  In order to describe this aggregation process, we employ master equations.  They define changes in the number of aggregates of a certain size in terms of reaction rates.  These reaction rates indicate the likelihood that two aggregates combine to form a large one, or that a large aggregate splits into two smaller parts.  The reaction rates are obtained from the simulations for a range of temperatures.

Our results indicate that the rates are not only temperature dependent, but also a function of the sizes of the aggregates involved in the reaction.  Using the measured rates, solutions to the master equations are shown to be stable and in agreement with the aggregate size distribution, as obtained directly from simulation data.  Furthermore, we show how temperature induced variations in these rates give rise to the observed changes in the aggregate distribution that characterizes the sol-gel transition.
\end{abstract}

\title{The Aggregation Kinetics of a Simulated Telechelic Polymer}%
\author{Mark Wilson}%
\affiliation{Department of Computational Science, San Diego State University, San Diego, California, USA}
\author{Avinoam Rabinovitch}%
\affiliation{Department of Physics, Ben-Gurion University of the Negev, Beer-Sheva, Israel}
\author{Arlette R. C. Baljon}%
\affiliation{Department of Physics, San Diego State University, San Diego, California, USA}
\date{\today}%
\maketitle

\section{INTRODUCTION}
\label{sec1}

Applications of polymer gels range from traditional thickeners for paint \cite{Glass} to novel precision gels for medical applications \cite{Werten}.  In all cases, their ability to form reversible cross-links plays a crucial role since it allows the system to remodel the internal structure under the application of external stresses.  Often these gels consist of triblock copolymers, containing a long hydrophilic spacer and short hydrophobic end groups \cite{Werten, Ma, Kujawa, Shen}.  In an aqueous solution, these end groups aggregate.  At high enough concentrations, aggregates form junction points in a gel network.  At lower concentrations, the hydrophobic groups still aggregate, but separate micelles form.  The sol-gel transition characterizes the cross-over from fluid like to solid like behavior.  Besides concentration changes both lowering the temperature or the application of external stress can also induce such a cross-over.  A unique feature of the material is that the junction points are not permanent, instead the aggregate size -  as well as its location - changes over time.

Understanding the structure of the gel network as well as its dynamics due to formation and breakup of aggregating end blocks is crucial to the development of optimal designed materials.  To this end, we have constructed a computer simulation of a toy system of associating telechelic polymers, in which end groups can form aggregates.  Each polymer chain, 8 units in length, is modeled as a bead-spring chain molecule \cite{Kremer}.  By means of a Monte Carlo step, we allow a chain end group to bind chemically with another end and hence form reversible junctions.  Aggregates contain end groups that are connected to each other by these junctions.  Their size varies and is set by many factors, such as the diffusion rate, the temperature of the system, and steric constraints due to the connection of each end group to a chain molecule.  In previous work, we have calculated the aggregate size distribution for a range of temperatures \cite{Baljon_Flynn} and characterized the sol-gel transition.  Moreover, we studied the network topology in detail using graph theoretical concepts \cite{Billen}.  In this work, we report on the kinetics of the aggregation process and analyze the results in terms of the reaction rates in the master equations. 

Several other groups have studied the sol-gel transition by means of computer simulations.  However, as far as we are aware, all of them have concentrated on the changes in structural properties instead of the kinetics underlying the reversible aggregation processes.  For instance Bedrov \textit{et. al.} \cite{Bedrov} performed simulations in which chain end groups were given a greater affinity to each other than to all other groups.  They find similar aggregate size distributions as in our work, however since there are no specific interactions that bind end groups together, reaction rates cannot be determined in this model.  The same is true for a recent study by Pandmanabhan \textit{et. al.} in which they investigate semi flexible polymer gels \cite{Pandmanabhan}.

This paper is organized as follows.  In section \ref{sec2}, we give the background of the simulation model, which is followed by a discussion of the master equations and reaction rates in section \ref{sec3}.  In section \ref{sec4}, we report on the measured rates and show that they are consistent with the aggregate distributions.  Section \ref{sec5} discusses how changes with temperature in the reaction rates are related to those in the overall dynamics of the systems that characterize the sol-gel transition.  This is followed by a conclusion in section \ref{sec6}.

\section{SIMULATION}
\label{sec2}

We employ a hybrid molecular dynamic (MD) / Monte Carlo (MC) simulation of telechelic polymers.  The simulation uses a course-grained model introduced by Kremer and Grest \cite{Kremer}.  Polymers are modeled as a string of beads. Beads interact with each other through a Lennard-Jones (LJ) potential
\begin{equation}
	U^{LJ}_{ij} = 4\varepsilon \left[ \left( \dfrac{\sigma}{r_{ij}}\right)^{12} -\left( \dfrac{\sigma}{r_{ij}}\right)^{6} - \left( \dfrac{\sigma}{r_{c}}\right)^{12} + \left( \dfrac{\sigma}{r_{c}}\right)^{6}\right],
\end{equation}
for $r<r_c$ and zero otherwise.  To speed up the calculations the cut-off distance was taken to be $r_c = 2^{1/6}$ and hence only the repulsive component of the potential is contributing. Beads connected by the chain structure are bonded through an anharmonic finitely extensible nonlinear elastic (FENE) potential  
\begin{equation}
	 U_{ij}^{FENE} = -\dfrac{1}{2}\kappa R_0^2 \,\, ln \left[ 1-\left( \dfrac{r_{ij}}{R_0} \right)^2\right],
\end{equation}
for $r_{ij} < R_0$ and infinite otherwise. The parameters used were $R_0 = 1.5\sigma$ and $\kappa = 30 \varepsilon \sigma ^{-2}$. It has been observed within the simulations \cite{Baljon_Flynn} that the average length of a FENE bond is $0.97\sigma$ and that the chains cannot cross through each other. The positions of the beads are updated in MD simulations by integrating the equations of motion using a 5th-order Gear predictor-corrector algorithm with a time step of $0.005\tau$.  Length, energy, and time are expressed in terms of the parameters ($\sigma, \varepsilon, \tau$) of the LJ potential.

The simulation system is comprised of 1000 telechelic polymer chains, each chain being eight beads in length.  The simulation cell has dimensions $24 \times 21 \times 27$ and hence the density is far below that at which the system turns glassy due to caging effects \cite{Baljon_Flynn}.  Periodic boundary conditions are employed in the first two directions, while solid surfaces confine the simulation in the third direction.  The beads at the chain ends are functionalized linkers and can form reversible junctions with multiple other end groups.  Formation and breaking of junctions takes place through MC moves \cite{Baljon_Vorselaars}.  Junctions are modeled by a FENE potential with the same parameters as described earlier, to which a constant negative constant $U_{assoc}$ is added.  Every $0.1\tau$, an attempt is made to form and break junctions between all end groups. For the probability of a formation success a Metropolis algorithm is used which depends upon a Boltzmann factor $exp(-\Delta U/kT)$, where $\Delta U = U^{FENE} + U_{assoc}$ is the energy difference between the old and the potentially new state.  The magnitude of $U_{assoc}$ affects the overall dynamics within the simulation and hence the temperature at which gelation occurs.  As in previous work \cite{Baljon_Flynn}, we choose $U_{assoc} = -22$, which implies that the micelle transition temperature is $T_m = 0.5$.   Note that this algorithm employs a simplification of neglecting the energy barrier $E_B$, between the bonded and the unbonded states, that is present in an experimental system.  It assumes a two-state system in which chain ends are either bound or unbound.  As has been pointed out by Hoy \textit{et. al.} \cite{Hoy}, the effect of $E_B$ would have been to slow down the rate of formation and breakage of individual junctions by a factor $exp(-E_B/kT)$.  However the equilibrium constant, defined as the ratio of these rates, is independent of barrier height.  Hence equilibrium properties such as aggregate size distributions are not influenced by our simplification.  It should be kept in mind that all rates at which aggregates form or break, reported later in this paper, are dependent on the chosen form of the interaction potential between the chain end groups (a simple FENE bond to which a constant negative association energy is added).  They are also dependent on the fact that we attempt to break and from junctions every $0.1 \tau$.  In this study we take these settings as given and aim to relate the structural changes observed when cooling the system to measured changes in formation and breaking rates. Since most of our results depend only on the values of the ratio of rates and hence are independent of the barrier, we believe this study to be relevant to experimental systems as well.
 
Finally we emphasize, that in our model there is no limitation to the number of junctions that an end group can form, hence detailed balance is observed within our work.  Hoy \textit{et. al.} \cite{Hoy} adapted the model to perform a study involving only binary interactions, and therefore in that case strict detailed balance was lost.

The simulation is analyzed through a range of temperatures by thermally cooling from an initially high temperature.  The temperature is controlled by coupling the simulation cell to a heat bath according to the fluctuation dissipation theorem as described by Kremer \textit{et. al.} \cite{Kremer}.  At each desired temperature, the system is allowed to equilibrate for at least $5000\tau$ prior to acquiring data.  The spatial locations of each bead within the system, along with end group junction data is then gathered for a number of time steps necessary achieve statistically consistent results.  Data at lower temperatures are obtained starting from a well-equilibrated configuration at $T = 1.5$. At this temperature, the system is first run with $U_{assoc} = 0$, so no junctions form.  Given that each chain is only eight beads long, equilibrium is reached quickly.  Junctions between chain ends are then introduced and the system is equilibrated until the number of junctions and the size distribution of the clusters of end groups fluctuates around its equilibrium value.  The system is then slowly cooled at a rate of $2.500\tau$ per $\Delta T = -0.1$, in order to reach the desired temperatures. Data obtained by cooling the system from different initial states at $T = 1.5$ are observed to be identical within statistical fluctuations.

At the temperatures considered in this study the system is quite mobile.  For instance the diffusion rate of aggregates equals $3\times 10^{-4} \, 1/\tau$ at $T = 0.45$ and approximately $10^{-2}\, 1/\tau$ at $T = 1.0$.  A single junction has an average lifetime of approximately $10^3 \tau$ at $T = 0.45$. The reported data have been obtained from runs that are at least ten times the inverse diffusion rate of the system; hence the system samples many different configurations during this computer experiment.  

\begin{figure}[htb]
	\includegraphics[width=\columnwidth]{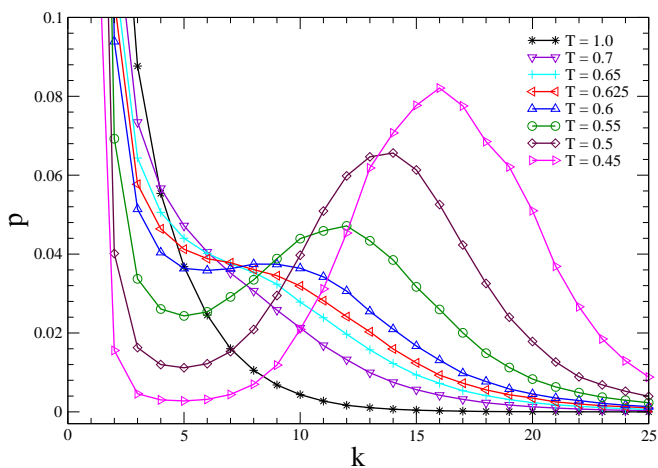}	
	\caption{\textbf{(Color online)} The time averaged aggregate probability distribution $p_k$ of the polymer system for a range of temperatures.\label{fig1}}
\end{figure}

  The aggregate size distribution is displayed in Fig \ref{fig1}. $p_k$ is the probability of finding an end group in an aggregate of size $k$, where an aggregate contains only end groups that are connected to each other through the reversible junctions.  It is calculated as:   
  \begin{equation}
 		p_k = \dfrac{N_k}{\sum_{k}N_k} = \dfrac{N_k}{N_{tot}},
 		\label{eq3}
 	\end{equation}	
where $N_k$ is the average number of aggregates of size $k$ and is obtained from the simulation data.  $N_{tot}$, the average total number of aggregates, depends on the temperature.

At high temperatures $p_k$ decreases monotonically with increasing cluster size.  However, at $T = 0.6$ a shoulder has formed, which develops in a pronounced peak at lower temperatures.  The temperature at which the shoulder first forms (approximately $T = 0.61$) is called the onset temperature \cite{Baljon_Flynn, Bedrov, Dudowicz}.  At this temperature the properties of the system first start to deviate from those of an ordinary liquid.  For instance, we have shown in previous work \cite{Baljon_Flynn} that below this temperature the dependence of the relaxation time on temperature is not any longer Arrhenius.  When cooled even further the system undergoes a micelle transition.  The total number of junctions between end beads $\Phi$ grows sharply and as a result a finite preferred aggregate size exists below the micelle transition.  At this point, the distribution $p_k$ exhibits a distinct peak.  The micelle transition temperature is defined as the point at which ${\partial^2 \Phi} /   {\partial T^2} = 0$ \cite{Lin}. The specific heat peaks as well.  For the system at hand, we have calculated \cite{Baljon_Flynn} that the micelle transition temperature $T_m = 0.51$.  The micelle transition is an equilibrium transition in the systems structural properties.  In addition at the gel transition the system becomes dynamically arrested due to the long times that the end groups reside within specific aggregates.  This nonequilibrium transition in the dynamics of the system occurs in our simulations at $T = 0.4$. 

In this paper, we will investigate by how far these transitions in the properties of a gel forming system can be explained by the reaction rates at which aggregates of end groups combine and break up.

\section{KINETIC MODEL}
\label{sec3}

Reversible aggregation processes can be described as two competing effects: particle aggregation and cluster fragmentation.  The traditional starting point \cite{VanKampen} for treating aggregation is a set of equations that describe how the sizes of the aggregated clusters change over time, called the master equations.  These equations contain the rates at which aggregates form and break apart.  In our simulations, an aggregate forms when two smaller aggregates closely approach each other and a new junction between chain end groups connects them.  It breaks apart when all the junctions between two subunits of an aggregate disconnect.  

\begin{figure}[htb]
	\includegraphics[width=\columnwidth]{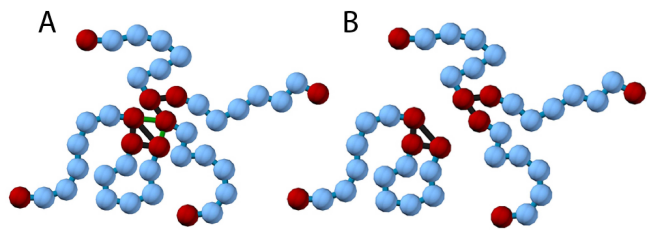}	
	\caption{\textbf{(Color online)} A 2D cartoon illustrating the breaking of an aggregate of size six \textbf{(A)} into two aggregates of size three \textbf{(B)}.  Note all simulations are preformed in 3D.  \label{fig1-2}}
\end{figure}

As an example of this process, we refer to a cartoon describing the breaking of an aggregate.  At the center of Fig. \ref{fig1-2}A six end groups (shown as red beads) are joined through multiple junctions.  These end groups comprise an aggregate of size six.  During the Monte Carlo step within the simulation, the two green junctions are broken, resulting in two aggregates of size three (Fig. \ref{fig1-2}B). The number of junctions broken, along with their distribution within the aggregate are determined according to the rules of the Metropolis algorithm. Note that aggregation within the system is a reversible process, therefore the time reverse reaction will also occur.

If we denote an aggregate of size $k$ by $(k)$, a change in the size of this aggregate can occur by the following reactions
  \begin{eqnarray}
     (\ell) + (k-\ell) \xrightleftharpoons[q_b(k,\ell)]{q_f(k,\ell)}  (k),
 		\label{eq4}\\
 		(k + \ell) \xrightleftharpoons[q_f(k+\ell,\ell)]{q_b(k+\ell,\ell)}  (k) + (\ell).
 		\label{eq5}
 	\end{eqnarray}	
In Eq. \ref{eq4}, $q_f(k,\ell)$ is defined as the rate that a given aggregate of size $\ell$  combines with one of size $k-\ell$ to form an aggregate of size $k$ within a time $\tau$.  Similarly for the reverse reaction, $q_b(k,\ell)$ is defined as the rate that an aggregate of size $k$ breaks into an aggregate of size $\ell$ and one of size $k-\ell$.  Of course an aggregate of size $k$ is also formed when an aggregate of size $k+\ell$ breaks into one of size $k$ and one of size $\ell$.  The reverse reaction destroys an aggregate of size $k$.  Eq. \ref{eq5} describes the latter two events.  

The master equations describing the aggregation model can be written as 
  \begin{eqnarray}
 		\nonumber&\dfrac{dN_k}{dt}& = \sum_{\ell=1}^{k-1} q_f(k,\ell) N_\ell N_{k-\ell} - \sum_{\ell=1}^{k-1} q_b(k,\ell) N_k   \\ 
 		&+&\sum_{\ell=1}^{\infty} q_b(k+\ell,\ell) N_{k+\ell} - \sum_{\ell=1}^{\infty} q_f(k+\ell,\ell) N_k N_\ell. 
 		\label{eq6}
 	\end{eqnarray}	 
Here, $N_k$ is the number of aggregates of size $k$.  The master equations represent a system of coupled differential equations. The model is finite since there are only 2000 end groups in the simulated system. Due to steric effects the maximum observed aggregate size is even smaller.  We never observed aggregates that contained more than approximately 60 end groups.

\section{RESULTS}
\label{sec4}
\subsection{Rates of Reactions}
The rates, at which the reactions according to Eq. \ref{eq4} and Eq. \ref{eq5} occur, are determined directly from the MD simulations. This is achieved by tracking the frequency of size specific reactions in both $(k)$ and $(\ell)$.  For instance, a formation occurrence $O_f(k,\ell)$ is defined as an event leading to an incremental change in the number of aggregates of size $k$ in a period of time $0.1\tau$, caused by the combination of an aggregate of size $\ell$ and $k-\ell$. The time averaged reaction rates are then determined as the ratio of $O_f(k,\ell)$ to the number of aggregates which can produce $(k)$.  Hence, $q_f(k,\ell)$ is determined such that,
\begin{equation}	
	 q_f(k,\ell) = \dfrac{O_f(k,\ell)}{N_\ell N_{k-\ell}}.	
	 \label{eq6.5}
\end{equation}	 
The breaking rate $q_b(k,\ell)$ is determined using a similar procedure.

\begin{figure}[htb]
	\includegraphics[width=\columnwidth]{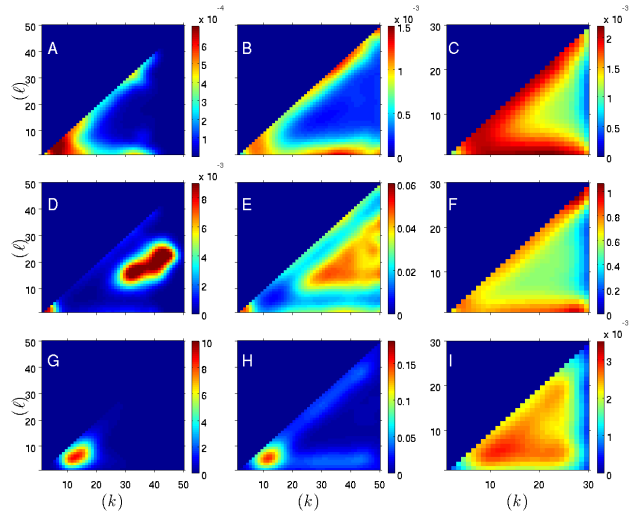}	
	\caption{\textbf{(Color online)} Average formation reaction rates $q_f(k,\ell)$ at a temperature of $T = 0.45$ \textbf{(A)}, $T = 0.55$ \textbf{(B)}, and $T = 1.0$ \textbf{(C)}.  The average breaking rates $q_b(k,\ell)$ are contained in \textbf{(D)}, \textbf{(E)}, and \textbf{(F)} respectively for each of the same temperatures.  The equilibrium constant $Q(k,\ell) = q_f(k,\ell) / q_b(k,\ell)$ are displayed in \textbf{(G)}, \textbf{(H)} and \textbf{(I)}.  Red indicates the most reactive $(k)$  and $(\ell)$, while dark blue is the least reactive.  Here, $(\ell)$ is assumed never to be greater than $(k)$.  Average rates have been smoothed for display purposes.\label{fig2}}
\end{figure}

	Fig. \ref{fig2} A, B, and C display color plots of the formation rates $q_f(k,\ell)$ at $T = 0.45$ (A), $T = 0.55$ (B), and $T = 1.0$ (C). Rates shown in red are most reactive, where as dark blue is the least reactive.  The breaking rates $q_b(k,\ell)$ are shown in (D), (E), and (F) for these temperatures respectively.  Within the determination of the rates, it is assumed that $(k)$ is always greater than $(\ell)$.  It is observed that the reaction rates of the polymer system are not only temperature dependent, but are also a function of the size of aggregates involved in the reaction. 
	
The equilibrium constant, defined to be the ratio of the reaction rates $Q(k,\ell) = q_f(k,\ell) / q_b(k,\ell)$, is related to the Gibbs free energy of the system and is shown at $T = 0.45$ (G), $T = 0.55$ (H), and $T = 1.0$ (I).  Regions of large magnitude $Q(k,\ell)$ mark $(k)$ and $(\ell)$ values, in which reactions are more favorable to occur.

\subsection{Consistency of the Master Equations with Aggregate Distribution}

\begin{figure}[htb]
	\includegraphics[width=\columnwidth]{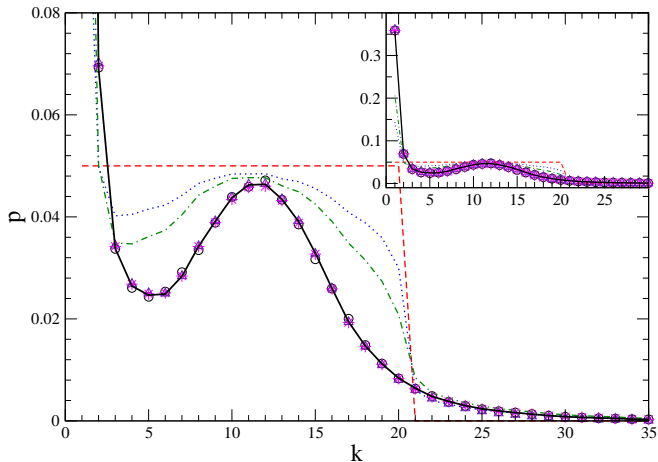}	
	\caption{\textbf{(Color online)} Solutions to the master equations of the kinetic model displaying agreement with the polymer aggregate distribution (circles) at $T = 0.55$.   The dashed and dotted lines show the time evolution of the master equations.  Starting from a constant (shown as dashed red), the distribution converges to the solid black line ($R^2 = 0.999$) after about $20\tau$.  Two intermediate steps are shown: the blue dotted line after $1\tau$ and the green dot-dashed one after $2\tau$. The triangles show that when only the $(\ell) = 1$ rates are employed the solution converges to the known aggregate distribution as well.  However, the convergence is slower and occurs after $500\tau$.   The convex optimization solution of Eqs. \ref{eq7} is shown as stars.  The inset is included so as to display the value of the solution at low $k$ values. \label{fig3}}
\end{figure}

The aggregate size distribution (Fig. \ref{fig1}) should result from the master equations (Eqs. \ref{eq6}) with the calculated reaction rates shown in Fig. \ref{fig2}.  Two different methods are used to prove that this is indeed the case.  The first consists of an unconstrained minimization of the master equations.  At equilibrium   ${dN_k} / {dt} = 0$ and the master equations become 
	 \begin{eqnarray}
 		0 &=& \sum_{\ell=1}^{k-1} q_f(k,\ell) N_\ell N_{k-\ell} - \sum_{\ell=1}^{k-1} q_b(k,\ell) N_k  \nonumber \\ 
 		&+&\sum_{\ell=1}^{\infty} q_b(k+\ell,\ell) N_{k+\ell} - \sum_{\ell=1}^{\infty} q_f(k+\ell,\ell) N_k N_\ell. 
 		\label{eq7}
 	\end{eqnarray}
Eqs. \ref{eq7} can be solved for $N_k$ using Newton-based convex minimization techniques \cite{Nocedal}.  Using Eq. \ref{eq3},  $p_k$ is obtained from the solution for $N_k$. The result at $T = 0.55$ is shown as stars in Fig. \ref{fig3}.  Note that the solution is in excellent agreement with the aggregate distribution obtained directly from the data (circles).

The second method consists of solving the master equations in time.  As an initial state, we take one in which each aggregate with $k \leq 20 $ is equally likely and no aggregates with size over 20 exist. The initial distribution is shown as a dashed red line in Fig. \ref{fig3}. This distribution is then updated by the master equations and the measured rates at $T = 0.55$ (the dotted blue line shows the distribution after time $1\tau$, and the dot-dashed green line after time $2\tau$). The aggregate distribution converges in approximately $20\tau$ to that shown as a solid black line.  As expected, this distribution agrees nicely with the measured aggregate size distribution (shown as circles).  Similar agreements were obtained for other initial distributions and other temperatures.

\subsection{Simplification}

The condition of detailed balance \cite{Chandler} can be used to show that only the rates of reactions, where one end group joins an aggregate or breaks away from it, completely determine the aggregate distribution.  This condition, if applicable, states that at a steady state the frequency of a specific reaction in the forward direction should be equivalent to that of the reverse reaction.  In our case, that implies  
\begin{equation}
 	q_f(k,\ell) N_\ell N_{k-\ell} = q_b(k,\ell) N_k
 	\label{eq8}
\end{equation}
for all $(k)$ and $(\ell)$, in which $k > \ell$.  It would therefore follow that 
\begin{equation}
 	N_k = \dfrac{q_f(k,\ell)}{q_b(k,\ell)} N_{k-\ell} N_\ell.
 	\label{eq9}
\end{equation}
This relationship at equilibrium has been verified within the simulations to hold (within statistical fluctuations) for multiple values of $\ell$, $k$, and $T$.

Eq. \ref{eq9} has two implications in regards to the kinetic model.  The first is that the first two terms and the last two terms in Eq. \ref{eq7} add up to zero independently.  The second is that the right hand side of Eq. \ref{eq7} adds up to zero for each value of $\ell$ independently. In other words, knowledge of $q_f(k,1)N_1 / q_b(k,1)$ is sufficient to calculate the aggregate size distribution by propagating Eq. \ref{eq9} for $(\ell)=1$: $N_k = \left[{q_f(k,1)N_1} / {q_b(k,1)}\right] N_{k-1}$.  The aggregate distribution  $p_k$ can then be obtained from Eq. \ref{eq3}.  Alternatively, starting from any initial condition, say the dashed red line in Fig. \ref{fig3}, and updating using only the $(\ell) = 1$ rates should yield the correct aggregate distribution. This procedure was carried out and indeed yielded the correct distribution, see Fig. \ref{fig3} (triangles). The convergence, however, is approximately a factor of 25 slower than the one attained when all rates are used. 

Although the $(\ell) = 1$ rates lead to the same correct distribution, the slow convergence here is due to the fact that they solely allow for a subset of the entire kinetics provided by the model using the complete range of rates.  We make use of the fact that the $(\ell) = 1$ rates by themselves yield the observed aggregate distribution (see below) when we discuss how changes in these reaction rates give rise to the observed changes in the aggregate distribution that characterizes a micelle transition.
\begin{figure}[htb]
	\includegraphics[width=\columnwidth]{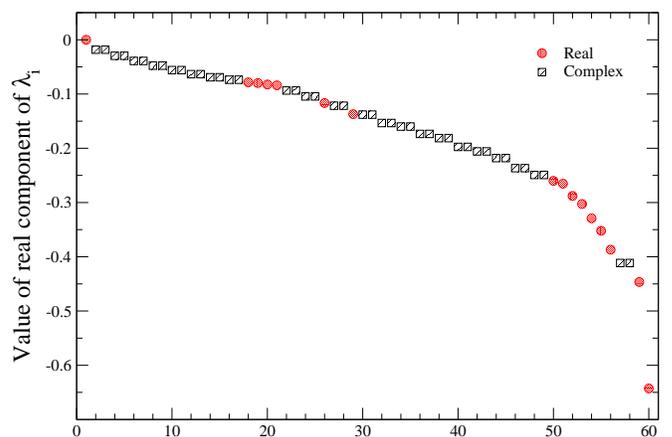}	
	\caption{\textbf{(Color online)} Magnitudes of the real component of the eigenvalues $\lambda_i$ of the master equations' Jacobian matrix, evaluated at the known equilibrium state of the polymer at $T = 0.55$.  The real eigenvalues are shown as red circles, while complex eigenvalues are shown in black squares. \label{fig4}}
\end{figure}
\subsection{Stability of Solutions}

To study the stability of the kinetic model under perturbations near the actual solution, we analyze the eigenvalues $\lambda_i$ of the Jacobian matrix \cite{Anosov}
\begin{equation}
 J =
\left[ {\begin{array}{cccc}
   \dfrac{\partial y_1}{\partial N_1} & \dfrac{\partial y_1}{\partial N_2} & \cdots & \dfrac{\partial y_1}{\partial N_m} \\
   \dfrac{\partial y_2}{\partial N_1} & \dfrac{\partial y_2}{\partial N_2} & \cdots & \dfrac{\partial y_2}{\partial N_m} \\
  \vdots  & \vdots  & \ddots & \vdots  \\
  \dfrac{\partial y_m}{\partial N_1} & \dfrac{\partial y_m}{\partial N_2} & \cdots & \dfrac{\partial y_m}{\partial N_m} \\
 \end{array} } \right],
\label{eq10}
\end{equation}
where $y_i$ represents the right hand side of $dN_i / dt$  according to Eq. \ref{eq6}.  The size $m$ of this matrix is determined such that continuous, non-zero valued rates exist for $1 \leq k \leq m$.  Above that value of $m$, the rates are undetermined within the simulation time. This procedure results e.g. in $m=60$ for $T = 0.55$.  Since reaction rates and the aggregate size distribution are known, the Jacobian can be evaluated for a given temperature, and its eigenvalues calculated. 

	The master equations consist of a multidimensional manifold. The resulting eigenvalues of Eq. \ref{eq10} are shown in Fig \ref{fig4}.  They are either complex with negative real parts (black squares) or real and negative (red circles).  This indicates that the solution is stable and that the singular point is a focus towards which the solutions are spiraling.  Both the smallest and largest eigenvalues are real.  The eigenvector corresponding to the largest absolute eigenvalue peaks at $N_1$.   Hence, the fastest process consists of establishing the correct number of singles. The eigenvector corresponding to the eigenvalue with the smallest absolute value is very similar to the equilibrium size distribution.  Apparently, the overall normalization of the solution is the slowest process \cite{Strogatz}.

\section{DISCUSSION}
\label{sec5}
\subsection{Rates of Reactions}

\begin{figure}[htb]
	\includegraphics[width=\columnwidth]{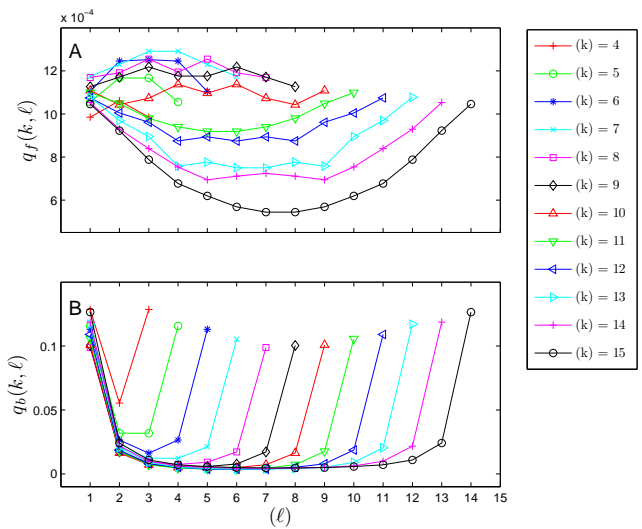}	
	\caption{\textbf{(Color online)} The average formation reaction rates $q_f(k,\ell)$ \textbf{(A)}, breaking reaction rates $q_b(k,\ell)$ \textbf{(B)} as a function of $(\ell)$ for several values of $(k)$ at a temperature of $T = 0.55$. Errors within the data are on the order of the symbol size. \label{fig5}}
\end{figure}

Fig. \ref{fig5}A displays $q_f(k,\ell)$ over a range of  $(k)$ values at $T = 0.55$.  A symmetry inversion with respect to $(\ell)$ is seen to exist within the formation reaction rates.  
For small $(k)$, the largest magnitudes of $q_f(k,\ell)$  are centered about $(\ell)=(k)/2$.  This fact indicates, for $(k)<11$, that two approximately equally sized aggregates participate in the formation reaction more frequently than an aggregate of small size $\ell$ and $k-\ell$.  The behavior changes with increasing $(k)$ and is displayed as an inversion of the symmetry within $q_f(k,\ell)$. For $(k)\geq 11$, where now the smallest magnitude of $q_f(k,\ell)$ is centered about $(\ell)=(k)/2$.  Here, aggregates of small size $\ell$ associate with $k-\ell$ sized aggregates more frequently than two approximately equal sized aggregates.  We label the $(k)$ value, at which the symmetry inversion of $q_f(k,\ell)$ occurs by $(k)_T$. At $T = 0.55$, $(k)_T=11$ and increases with decreasing temperature.  The disassociation reaction rate $q_b(k,\ell)$, shown in Fig. \ref{fig5}B, does not exhibit such symmetry inversion.  It is always easier to break off single beads than to split the system into equal parts.  Note that the rates in figure \ref{fig5}A and \ref{fig5}B are symmetric around $(k)/2$ since a split into aggregates of sizes $(\ell)$ and $(k-\ell)$ is identical to a split into sizes $(k- \ell)$ and $(\ell)$.

\begin{figure}[htb]
	\includegraphics[width=\columnwidth]{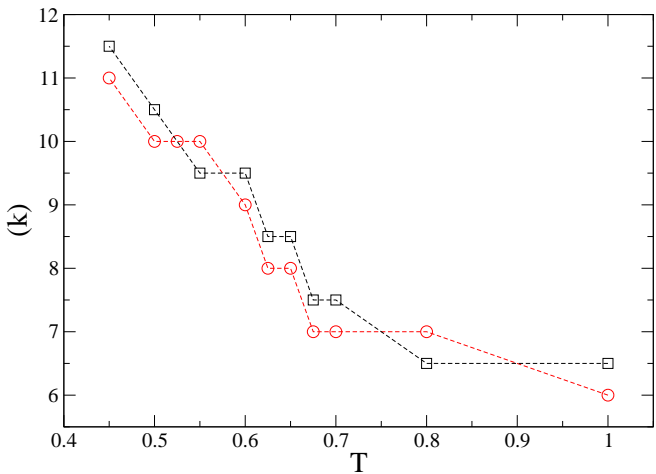}	
	\caption{\textbf{(Color online)} The most spherically symmetric aggregate size $(k)_S$ (red circles), and the aggregate size where in $q_f(k,\ell)$ exhibits an inversion in symmetry $(k)_T$ (black squares) as a function of temperature. Lines are shown as guides for the eye.\label{fig6-2}}
\end{figure}

Fig. \ref{fig6-2} compares $(k)_T$ values for a range of temperatures with the aggregate size $(k)_S$ at which the aggregates are most spherical. The sphericity is calculated from the eigenvalues of the gyration tensor as described in \cite{Kenward}. The smallest aggregates do not contain enough end groups to form a sphere.  On the other hand, the largest aggregates have a long wormlike structure since steric effects caused by the polymer chains prevent the formation of a very large sphere. Thus, the most spherical value appears at a middle aggregate size $(k)_S$.   As can be seen both $(k)$ values in Fig. \ref{fig6-2} match, indicating that if an aggregate has a size larger than $(k)_S$, it is most likely to grow by addition of single beads or small aggregates at one of its ends. 

\subsection{Dependence of Reaction Rates on Temperature}

\begin{figure}[htb]
	\includegraphics[width=\columnwidth]{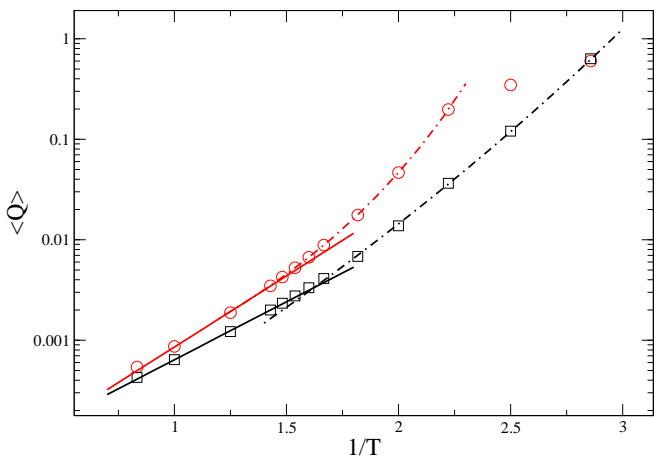}	
	\caption{\textbf{(Color online)} A semi-log plot of $\langle Q \rangle $ as a function of $1/T$, calculated for all $(k)$ and $(\ell)$ (red circles), and for $(\ell) = 1$  (black squares).  The solid lines display an Arrhenius fit to the data within the high temperatures, while the dashed lines correspond to a stretched exponential fit within low temperatures.   See the text for values of the fitting parameters. \label{fig6}}
\end{figure}

Fig. \ref{fig6} displays $\langle Q \rangle$ as a function of inverse temperature.  $\langle Q \rangle$ is calculated as a weighted average  
\begin{equation}
	\label{eq11}
 \langle Q \rangle = \dfrac{\sum_{k,\ell} Q(k,\ell) n(k,\ell)}{\sum_{k,\ell} n(k,\ell)},
\end{equation}
where $n(k,\ell)$ denotes the number of reactions involving aggregates of sizes $k$  and  $\ell$. Two separate cases are considered: (1) reactions in which only aggregates of size $\ell=1$ are involved, and (2) all reactions. 

	For high temperatures, Arrhenius behavior, as indicated by a linear trend on the semi-log scale, is observed.  The rate data is fit by an Arrhenius curve $\langle Q \rangle = \langle Q \rangle_0 exp (C/T)$  within the temperature range between $T = 1.2$ and 0.7  and is shown as a solid line.  In the case where $(\ell) = 1$ (shown as squares), the curve fit results in parameters of   $C = 2.65$ and $\langle Q \rangle_0 = 4.49\times10^{-5}$, whereas when all $(k)$ and $(\ell)$ rates are used (shown as circles), $C = 3.26$ and $\langle Q \rangle_0 = 3.29\times10^{-5}$.  The equilibrium constant deviates from being described by an Arrhenius curve as the temperature decreases from $T = 0.7$.  The low temperature data is fit by a stretched exponential $\langle Q \rangle = \langle Q \rangle_0 exp (C/(T-T_0))$  and is shown as a dashed line within the figure. The curve fitting results in parameters $T_0=0.084$, $C = 2.78$, and $\langle Q \rangle_0 = 1.8\times10^{-5}$ for the $(\ell)=1$ case, and $T_0=0.25$, $C=1.49$, and $\langle Q \rangle_0=1.3\times10^{-4}$ for the all rates data.  The extrapolated temperature of $T_0 = 0.25$ is in close agreement with our earlier published work \cite{Baljon_Flynn}, where the gelation temperature of $T_0=0.29$ for a similar system was determined from an analysis of relaxation times.  The two fits cross at $T = 0.65$ for the $(\ell)=1$ case and $T = 0.6$ when all $(k)$ and $(\ell)$ rates are used.  This is close to the ``onset'' temperature, as was observed in previous work \cite{Baljon_Flynn}. 
	
	Again, we note that the $(\ell)=1$ case describes a subset of the kinetics provided by the complete range of $q$'s within the model. At $T = 0.35$ within the figure, the two cases converge. This is to be expected, since it has been observed that at this temperature, reactions are primarily composed of a single end group joining an aggregate or breaking away from it.

\subsection{Relation Between Reaction Rates and Size Distribution}

\begin{figure}[htb]
	\includegraphics[width=\columnwidth]{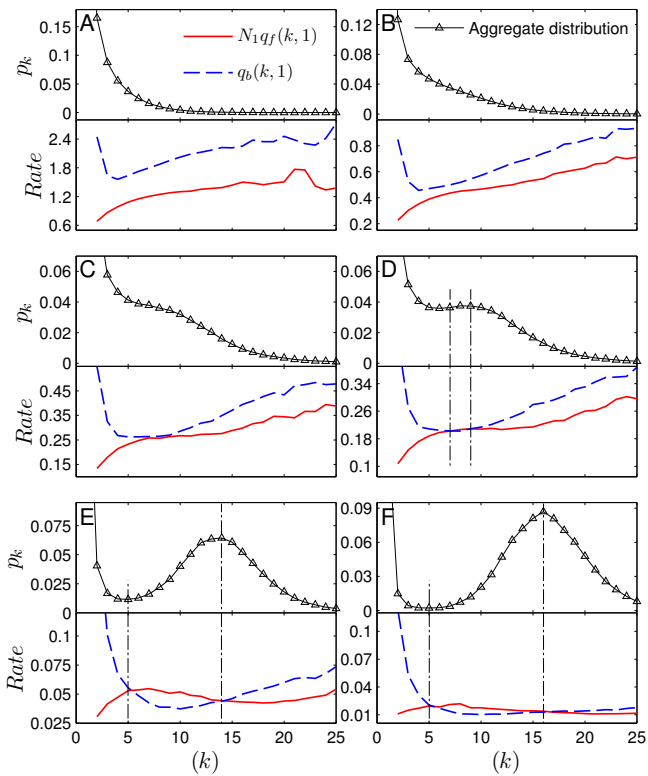}	
	\caption{\textbf{(Color online)} The probability distribution $p_k$ and the crossing of the rates $q_f(k,1)N_1$ with $q_b(k,1)$ at $T = 1.0$ \textbf{(A)},   $T = 0.7$ \textbf{(B)}, $T = 0.625$ \textbf{(C)}, $T = 0.6$ \textbf{(D)},  $T = 0.5$ \textbf{(E)}, and $T = 0.45$ \textbf{(F)}.   At $T = 0.625$, the rates nearly cross.  This temperature is in agreement with the onset of a shoulder within $p_k$.  For lower temperatures, the crossing of the rates is in agreement with the extrema of $p_k$. The dashed lines act as guides for the eye and mark the $(k)$ aggregate at which $p_k$ is at a maximum, in agreement with $(k)$ when $q_f(k,1)N_1=q_b(k,1)$.  \label{fig7}}
\end{figure}

\begin{figure}[htb]
	\includegraphics[width=\columnwidth]{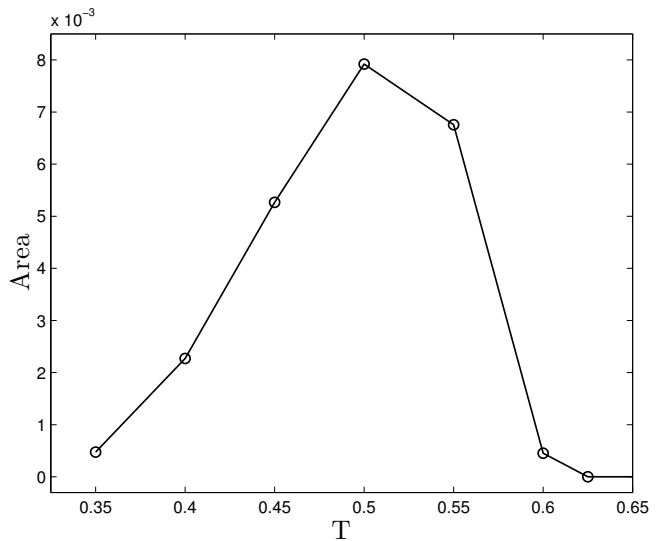}	
	\caption{The area calculated as $\sum_{k} \left[ q_f(k,1)N_1-q_b(k,1)\right] $ over the range of $(k)$ in which $q_f(k,1)N_1 > q_b(k,1)$  as observed in Fig. \ref{fig7} \textbf{(D)} through \textbf{(F)} as function of temperature.  \label{fig8}}
\end{figure}

One of the signatures of a micelle transition is the occurrence of a preferred finite aggregate size at low temperature \cite{Obeid} characterized by the peak in the aggregate size distribution.  We now investigate how this feature is related to the observed reaction rates. The only reactions considered are those in which one end group is added or removed from an aggregate.  As shown above the rates for these reactions completely determine the aggregate size distribution. Fig. \ref{fig7} shows the aggregate distributions together with these reaction rates for a range of temperatures. Note that $q_b(k,1)$ increases faster with $(k)$ than $q_f(k,1)N_1$.  
 With decreasing temperature, $q_f(k,1)N_1$ becomes less dependent on $(k)$ and ultimately exhibits a decreasing trend with increasing $(k)$. This feature is necessary for the two rates to cross and provides a range of $(k)$, where in $q_f(k,1)N_1 > q_b(k,1)$.  For these $(k)$ values, it is more favorable for a single end group to join the aggregate than to break from it. Moreover, at low temperatures, the minimum and maximum points in the aggregate size distribution are seen to correspond to the points where $q_f(k,1)N_1 = q_b(k,1)$.  This is theoretically easily verified.  An extremum in the aggregate distribution implies that $dp/dk = 0$ or $p_k\approx p_{k-1}$. Using Eq. \ref{eq3}, Eq. \ref{eq8} can be rewritten as
\begin{equation}	 
	\label{eq12}
	 q_f(k,\ell)N_{tot}\,p_\ell\,p_{k-\ell} = q_b(k,\ell)p_k, 
\end{equation}	 
or
\begin{equation}
	\label{eq13}	 
	 \dfrac{p_k}{p_{k-\ell}} = \dfrac{q_f(k,\ell)}{q_b(k,\ell)}N_{tot}\,p_\ell = \dfrac{q_f(k,\ell)}{q_b(k,\ell)}N_\ell, 
\end{equation}
for every $(\ell)$.  In particular, when $\ell=1$  
\begin{equation}
	\label{eq14}	 
	 \dfrac{p_k}{p_{k-\ell}} = \dfrac{q_f(k,1)}{q_b(k,1)}N_{tot}\,p_1 = \dfrac{q_f(k,1)}{q_b(k,1)}N_1.
\end{equation}
So $p_k=p_{k-1}$ when $q_f(k,1)N_1=q_b(k,1)$. At high temperature $q_f(k,1)N_1 < q_b(k,1)$ for all aggregates of size $k$ and hence the aggregate size distribution has no extreme.  The rates do not cross for $T = 0.625$ and higher, while they do for $T = 0.6$ or lower. The transition temperature $T \approx 0.61$ coincides with the temperature where the aggregate size distribution first forms a shoulder.  This is expected since a necessary condition for the appearance of a peak in the aggregate size distribution is that the curves cross.  Below $T = 0.61$, Arrhenius dependence of the equilibrium constants and relaxation rates on temperature is lost. Moreover, the area between the $q_f(k,1)N_1$ and $q_b(k,1)$ curves, shown in Fig \ref{fig8}, first increases and then decreases again when the temperature is lowered from $T = 0.6$.  The area has an extremum at $T = 0.5$: the micelle transition temperature.

Note that the above observation follows from the fact that both the forming and breaking rates are non monotonic as a function of $(k)$.  This fact deserves some further explanation.  Most importantly, the two rates depend differently on $(k)$, one being concave up and the other increasing or concave down.  We believe that these differences are due to cooperative effects.  In general a single bead is attached to an aggregate with more than one junction (see Fig. \ref{fig1-2}). In order for the bead to break away from the aggregate, multiple junctions have to break. The junctions have to break in a short time interval, since there is a significant chance that the first junction that broke would reform while the others are still there.  On the other hand, when a bead joins an aggregate it is sufficient that one junction forms.  $q_b(k,1)$ initially decreases with $(k)$ since at small $(k)$ only one junction has to be broken and the need for two or more to be broken increases with $(k)$.  In contrast, the larger the aggregate, the more places there are at which an end group can break off.  Hence, at some point the trend reverses and the rate starts to increase with $(k)$. On the other hand, $q_f(k,1)$ initially increases with (k).  This is due to the fact that a single bead can attach at more points to the aggregate of size $(k-1)$ as $(k)$ increases. For this reason, one would expect $q_f(k,1)$ to increase monotonically.  This is indeed the case at high temperature.  However at low temperature the rate starts to drop again at large $(k)$ values.  We believe that this is due to the much lower diffusion rate of large aggregates at low temperatures. 

\section{CONCLUSION}
\label{sec6}

Usually in solution and other dense media one has to check the validity of transition rate theory, since if friction is high, solvent dynamics cannot be neglected. This effect is very nicely demonstrated by Anna \textit{et.al.} \cite{Anna}, who showed experimentally how transition-rates change in different solvents, in good agreement with Kramers theory \cite{Kramers}. Kramers theory simulates solvent dynamics as frictional drag and influence of random forces. For the sake of simplicity we assumed here that both effects are small at the temperature range considered, and opted to use regular rate theory.

	We have investigated the kinetics of aggregation and fragmentation in a simulation of associating polymers.  The process was described in terms of master equations and rates of reactions were obtained from the simulation. The master equations were shown to have a solution consistent with the aggregate size distribution as obtained directly from the simulations.  Stability analysis showed that the aggregate distribution was indeed a stable solution to the master equations.  We found that, since detailed balance holds within the simulation, knowledge of the $(\ell) = 1$ rates provides all the necessary information to reproduce the complete aggregate size distribution. 

It was observed that changes with temperature of the reaction rates can explain the changes in the aggregate distribution and dynamics near the sol-gel transition.  The onset temperature of this transition was determined as the temperature at which there first exist a range of $(k)$ values for which it is more likely that a single end group attaches to an aggregate than that one breaks off ($ q_f(k,\ell)N_1 > q_b(k,\ell) $).  Near this temperature,   $\langle Q \rangle$ begins to deviate from the Arrhenius, fluid-like behavior.  In addition a shoulder develops within the distribution.  Below the onset temperature, aggregates have a preferred size.  A symmetry inversion within the formation rate $q_f(k,\ell)$ occurs at this size and marks a fundamental change in the aggregate formation process.  Aggregates smaller than the preferred size tend to be formed by joining two equal sized aggregates, whereas larger aggregates are formed by the addition of one end group.  As temperature continues to decrease, the likelihood that $q_f(k,\ell)N_1$ is larger than $q_b(k,\ell)$ (measured as the area between the two curves of the rates versus $(k)$) grows. The maximum at $T = 0.5$, is in agreement with the micelle transition temperature.  

In future work, we plan to investigate how the reaction rates are influenced by uniform and oscillatory shear. It is well known \cite{Oles} that the application of external stresses can influence the aggregate distribution; however its influence on the kinetic processes in the system has been studied less.  The master equations will be most likely a valuable tool in such studies as well.

\begin{acknowledgments}
This work is supported by the National Science foundation under DMR-1006980 and CHE-0947087.
\end{acknowledgments}


%

\end{document}